\newif\ifpub \pubtrue

\ifpub
\documentclass{elsart}
\else
\documentclass[twocolumn]{elsart}
\fi

\usepackage[dvips]{graphicx} 
\usepackage{amssymb} 
\usepackage{threeparttable}
\usepackage{url} 
\usepackage{mcite} 

\newcommand{\hs}{\hspace{0.5in}}

\newcommand{\tGasStuffCaption}{
Physical parameters of $^8$Li gas transport system. Ranges indicate approximate
working values during operation, and are in rough agreement with the
gas model given in the text.}

\newcommand{\tGasStuff}{
\begin{table*}
\begin{center}
\ifpub  \caption{}{}
\else   \caption{}{\tGasStuffCaption}
\fi
\begin{tabular}{|l|l|}
\hline \hline
Production (target) Chamber &\\
\hs	Volume (without collar)  	& 510 cm$^3$ \\
\hs	Volume (with collar)		& 279 cm$^3$ \\
\hs	Operating pressure (absolute)	& 195-210 kPa \\
\hline
Conduit transfer line &\\
\hs	Material 	&	Teflon \\
\hs 	Inner diameter	&	3.18 mm\\
\hs	Length		&	40 m  \\	
\hline
Umbilical &\\
\hs     Inner capillary material	& PTFE \\
\hs	Inner capillary inner diameter 	& 3.38 mm \\
\hs	Inner capillary outer diameter	& 4.14 mm \\
\hs	Exhaust tube outer diameter	& 8.0 mm \\
\hs   	Overall diameter with sheath	& 15.8 mm \\
\hs	Length				& 26 m \\
\hline
Decay Chamber &\\
\hs     Radius                          & 6.35 cm \\
\hs	Volume				& 1070 cm$^3$ \\
\hs	Internal operating pressure (absolute)	& $\sim$125 kPa \\
\hline
Transport Gas &\\
\hs Composition          		& He ( $\sim 0.1 \%$ N$_2$), \\
					& plus NaCl aerosol\\
\hs Viscosity \cite{CRC}		& 2.0 std. atm. s  \\
\hs Mass Flow Rate (Q)			& 250-300 std. atm. cm$^3$/s \\
\hline
\end{tabular}

\label{t:gas_stuff}
\end{center}
\end{table*}
}

\newcommand{\myfig}[3]{
\ifpub  
\begin{figure}[p]
\begin{center}
\includegraphics[width=\textwidth,height=\textheight,keepaspectratio=true]{#2}
\vspace{1in}
\caption{}{}
\label{#1}
\end{center}
\end{figure}
\clearpage
\else
\begin{figure*}
\begin{center}
\includegraphics[width=0.8\textwidth,height=\textheight,keepaspectratio=true]{#2}
\caption{}{#3}
\label{#1}
\end{center}
\end{figure*}
\fi
}

\newcommand{\fToidecayCaption}{
The $^8$Li and $^8$B decay schemes. Data taken from Reference \cite{toi}.
} 
\newcommand{\fToidecay}{\myfig{f:toidecay}{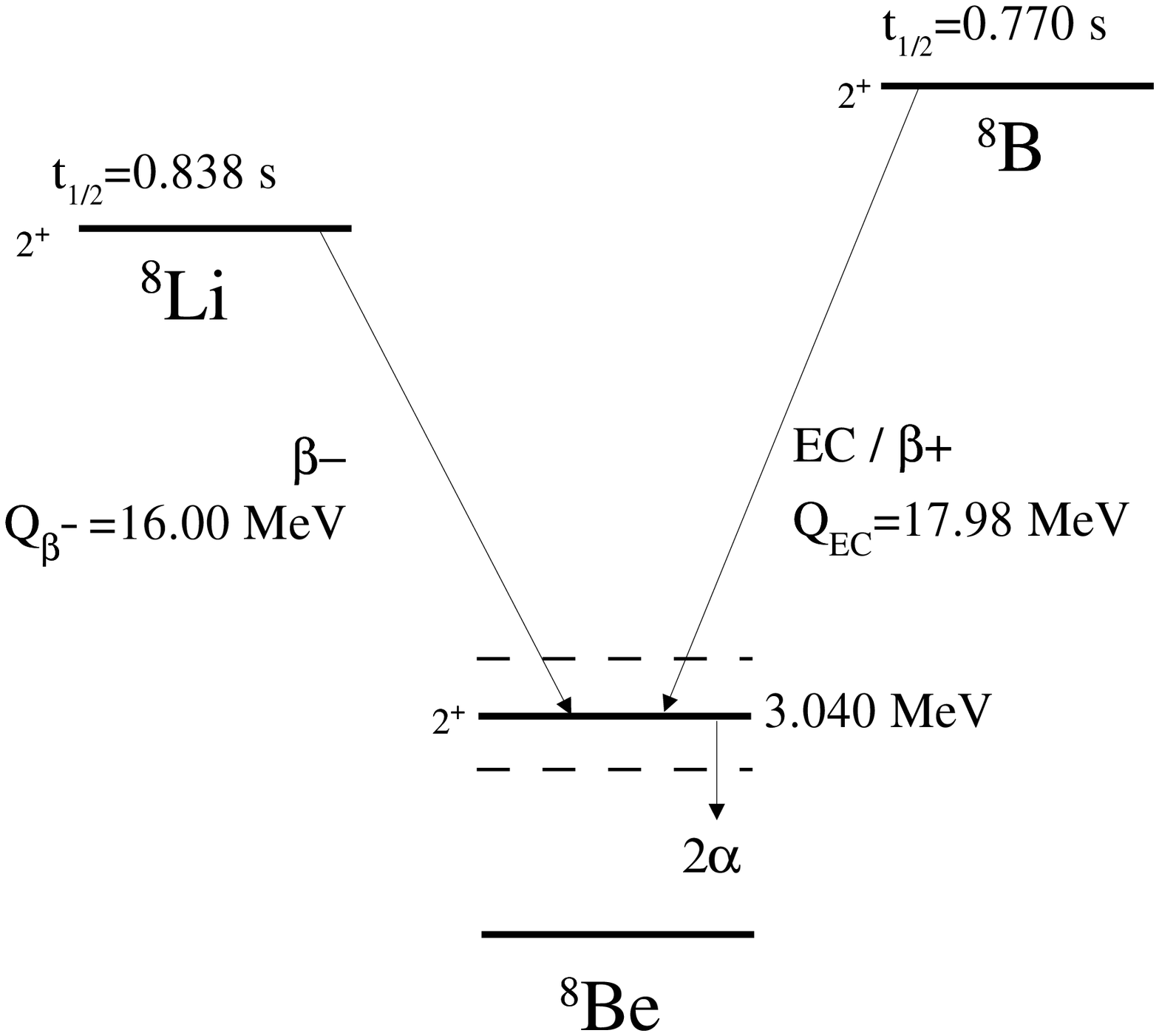}{\fToidecayCaption}}

\newcommand{\fOverviewCaption}{
The basic features of the $^8$Li calibration source.  Radioisotope is
produced using fast neutrons impingent on a $^{11}$B-enriched
target. Helium gas containing NaCl aerosol is used as a transport
medium to carry $^8$Li from the target chamber to the decay chamber,
where $^8$Li decays give off $\beta$-particles, which escape into the
SNO detector volume, and $\alpha$-particles, which are tagged within the
decay chamber.  }
\newcommand{\fOverview}{\myfig{f:overview}{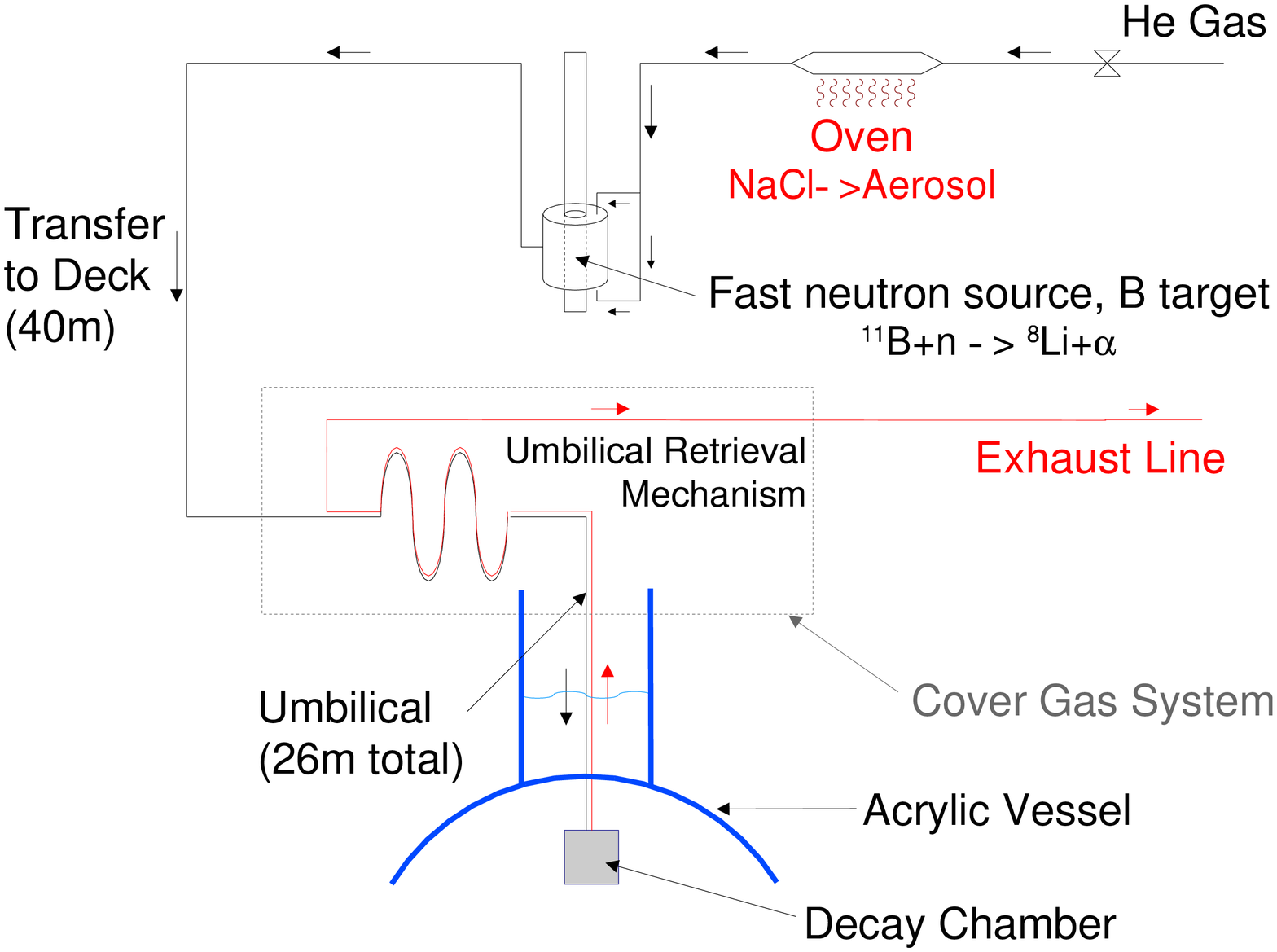}{\fOverviewCaption}}

\newcommand{\fTargetChamberCaption}{
The target (production) chamber, {\it in situ} about the neutron
source. Gas flows into the volume through holes in perforated
distribution plates. A thin cylinder of $^{11}$B-enriched boron in polyethylene
sits inside the gas volume, and a thin layer of boron nitride coats
the ouside of the gas volume. $^8$Li atoms recoil into the gas volume
and are swept by laminar flow to a series of holes around the
circumference of the chamber, which connect to the output line.  }
\newcommand{\fTargetChamber}{\myfig{f:target_chamber}{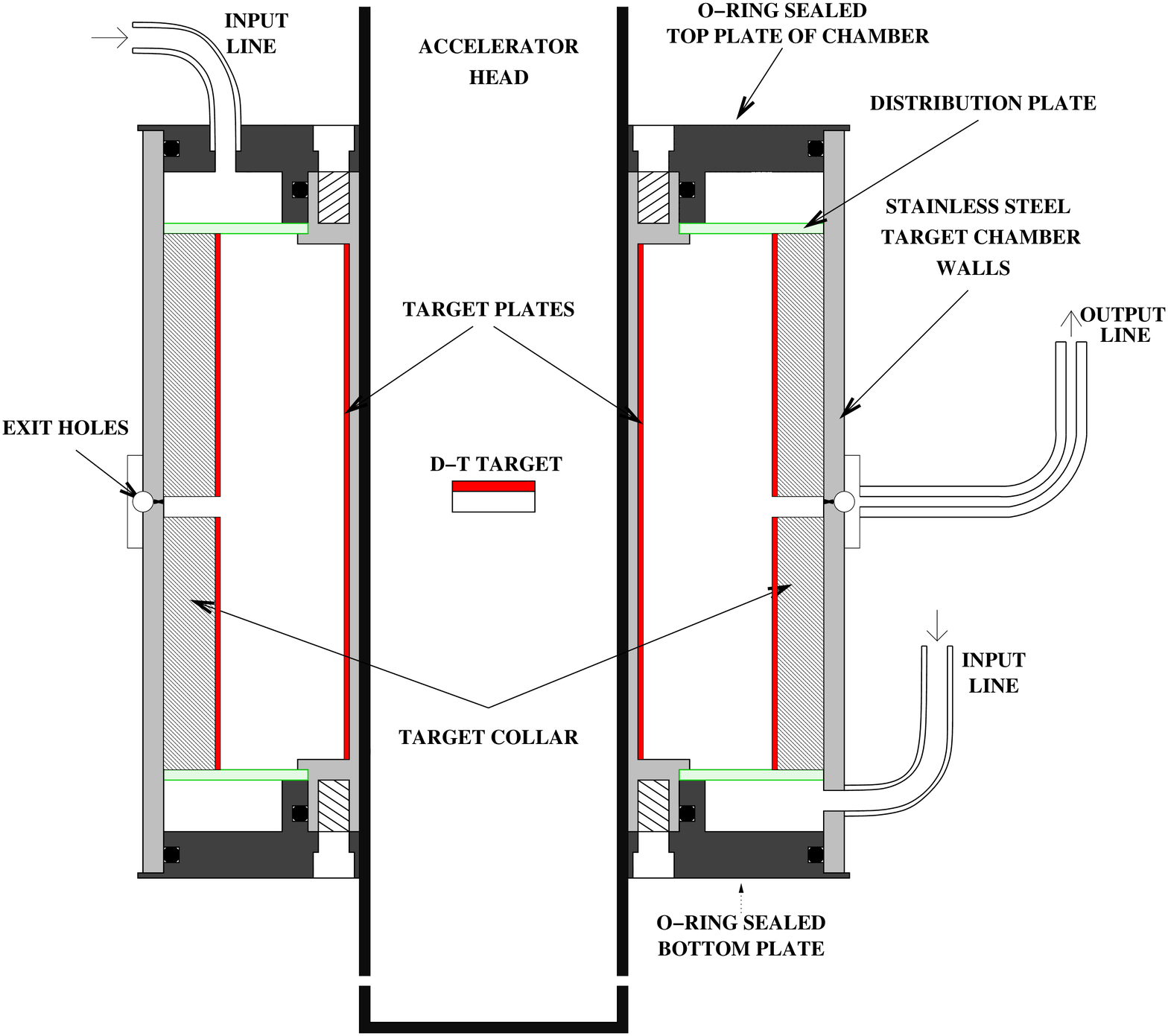}{\fTargetChamberCaption}}

\newcommand{\fEfficiencyVsDiameterCaption}{
Partial efficiencies for exiting the target chamber, transport across
the capillaries, and the efficiency for the decay in the trigger
chamber are shown for different transport tube diameters. Also given is the overall efficiency.}
\newcommand{\fEfficiencyVsDiameter}{\myfig{f:efficiency_vs_diameter}{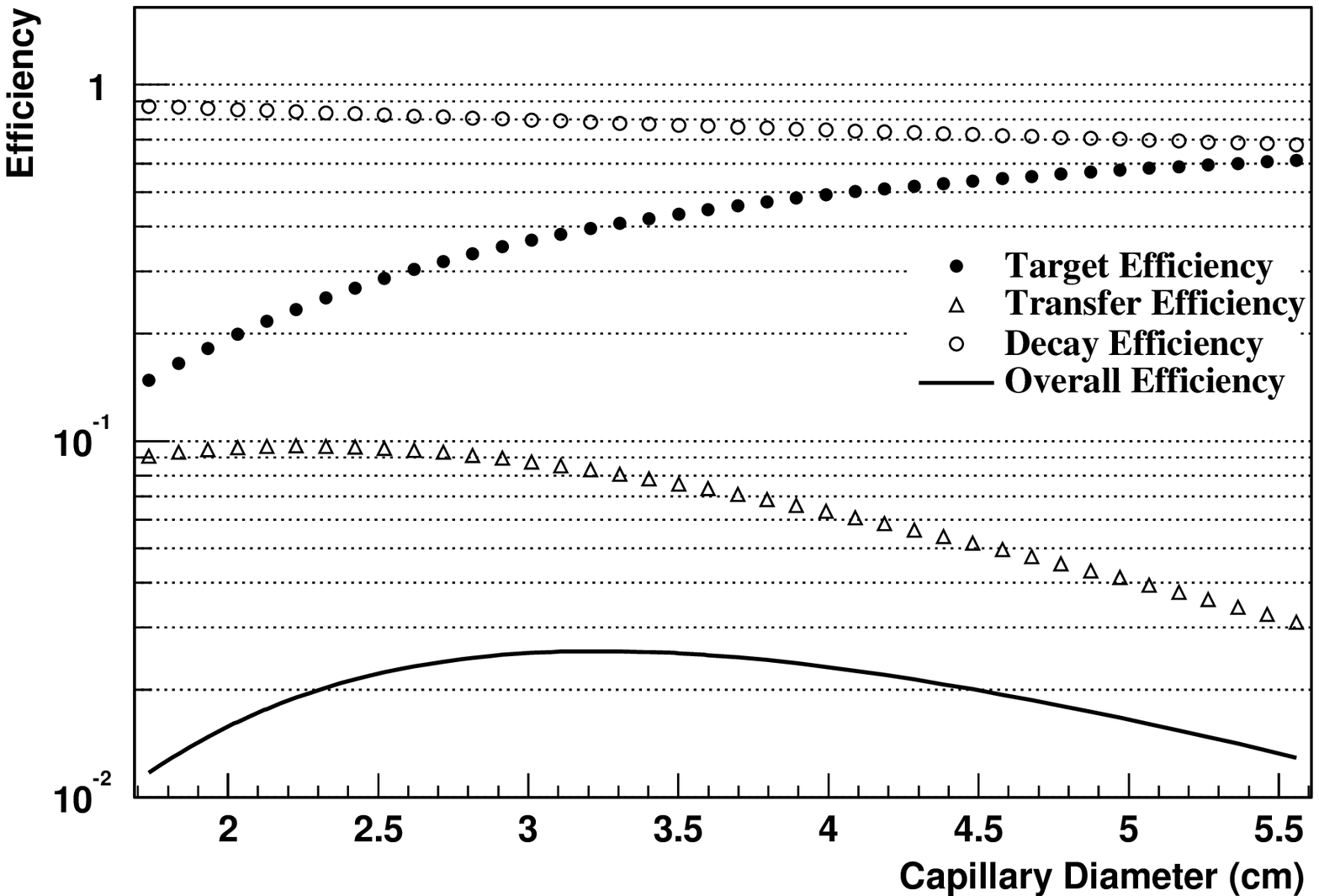}{\fEfficiencyVsDiameterCaption}}

\newcommand{\fSourceCaption}{
The $^8$Li decay chamber. Two thin hemispheres of
spun stainless steel are electron-welded together to create the decay
volume at the bottom.  These are in turn welded to the ``neck''
which holds the gas fixtures, and has a central hole for a 4.86~cm
diameter acrylic window. Atop this window is a 3.81~cm diameter PMT.  The
PMT and gas fittings are contained within a sealed stainless steel
endcap. Gas flows both in and out near the top of the volume.  The top of this endcap holds the end of the umbilical
(sealed with compressed O-rings). The source is supported from above by the SNO
source manipulator system.  Measurements are given in centimeters.
}
\newcommand{\fSource}{\myfig{f:source}{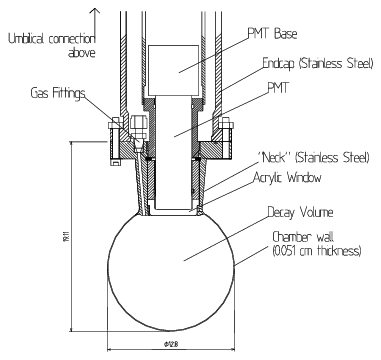}{\fSourceCaption}}

\newcommand{\fTagSignalCaption}{
The integrated charge and time of signals coming from the decay
chamber PMT for 47823 tagged events. Contaminant $\beta$-particles are
clearly distinguishable from the $\alpha$ scintillation signal from
$^{8}$Li. The line shows the cut used to discriminate against
contamination events.  }
\newcommand{\fTagSignal}{\myfig{f:tag_signal}{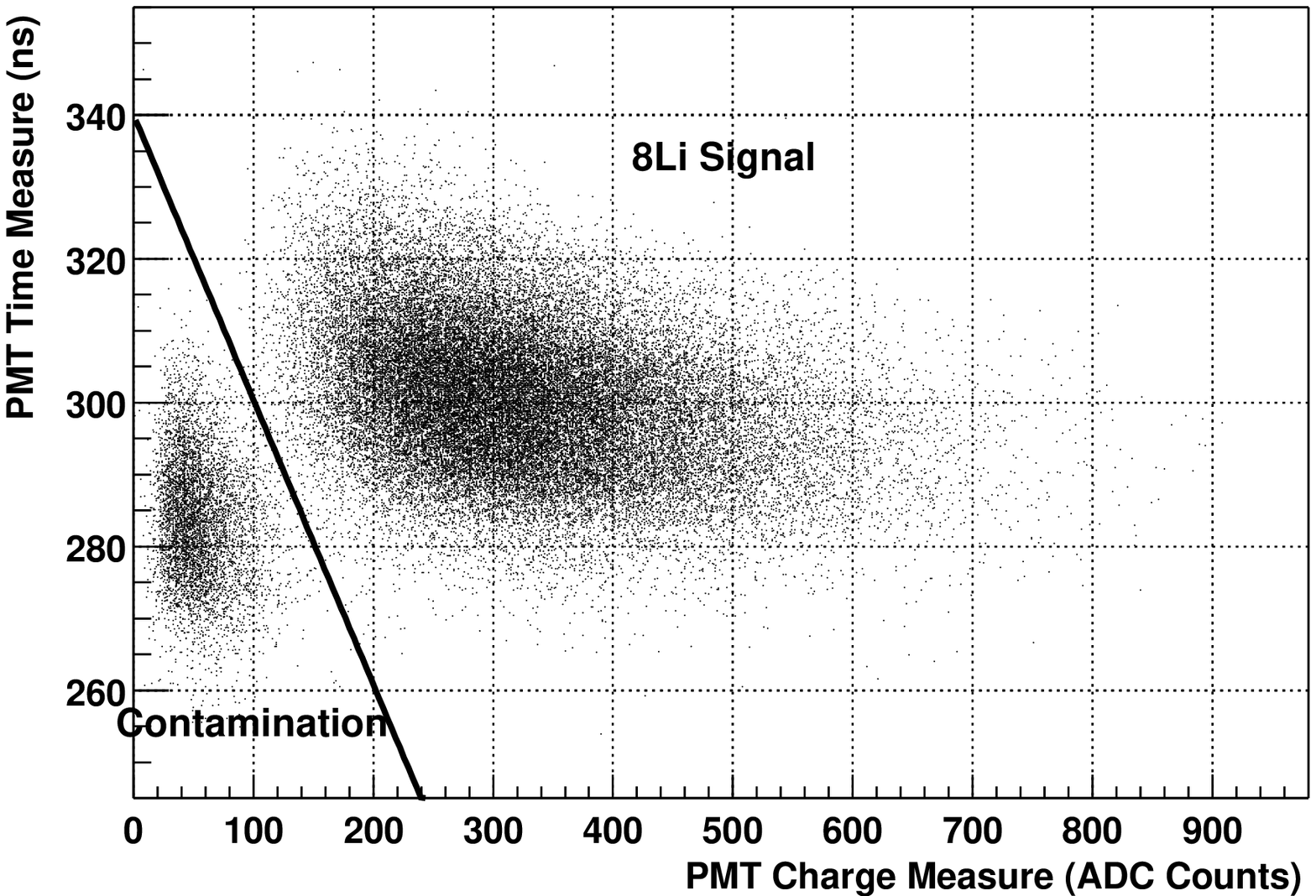}{\fTagSignalCaption}}

\newcommand{\fRawNhitCaption}{
The N$_{hit}$ spectra are shown for all source-related events within
the detector during $^{8}$Li runs. (Note this spectrum is dominated by
the $^{16}$N gamma contamination described in the text.)  Also shown
are spectra after cuts on the decay chamber PMT signal designed to
distinguish tagged contaminant events from the tagged $^{8}$Li
events.}
\newcommand{\fRawNhit}{\myfig{f:raw_nhit}{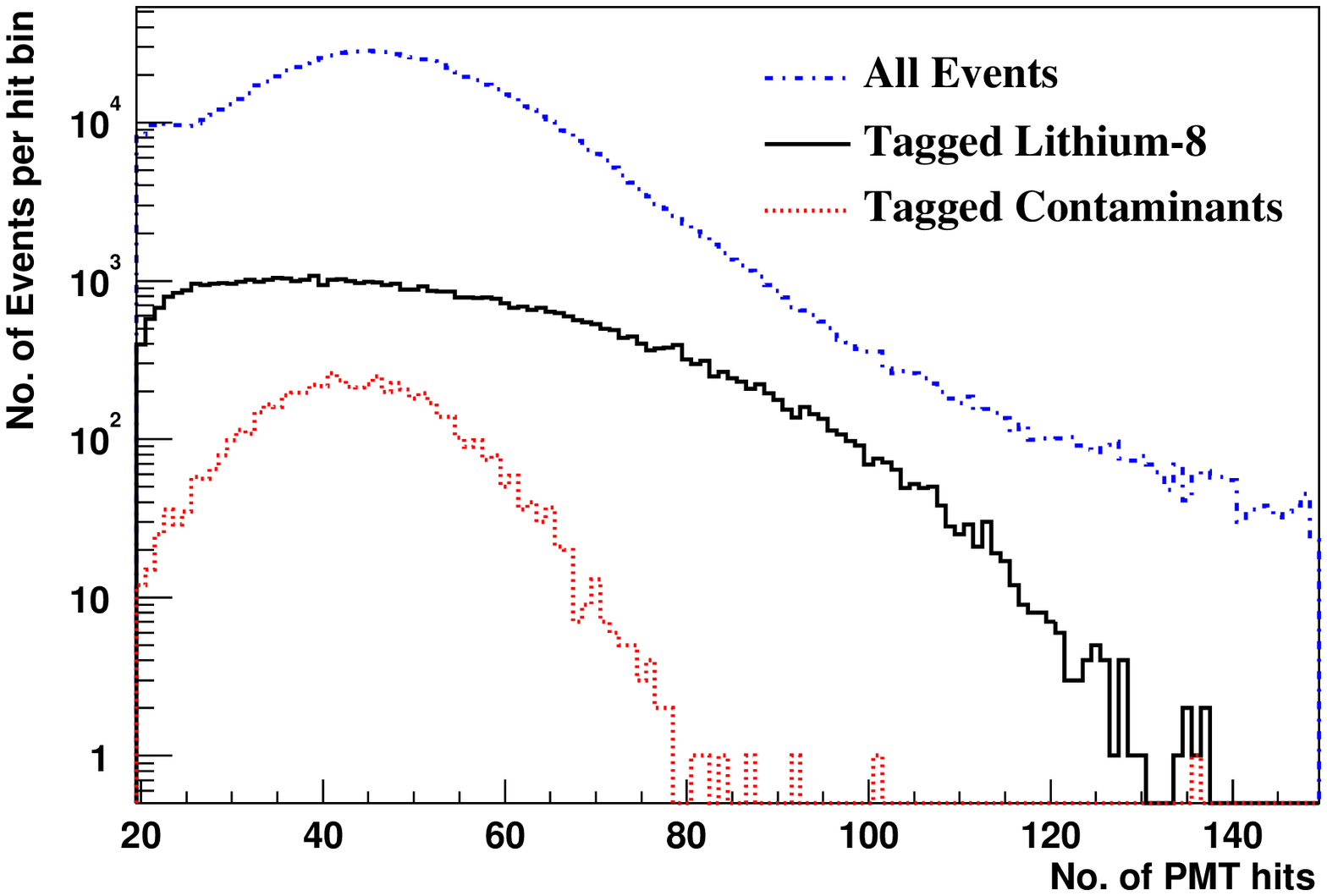}{\fRawNhitCaption}}

\newcommand{\fLiNhitCaption}{
The N$_{hit}$ spectrum of tagged $^8$Li events is
shown compared to a SNO Monte Carlo prediction of the source. The
Monte Carlo energy scale was tuned using the $^{16}$N calibration
source.
}
\newcommand{\fLiNhit}{\myfig{f:li_nhit}{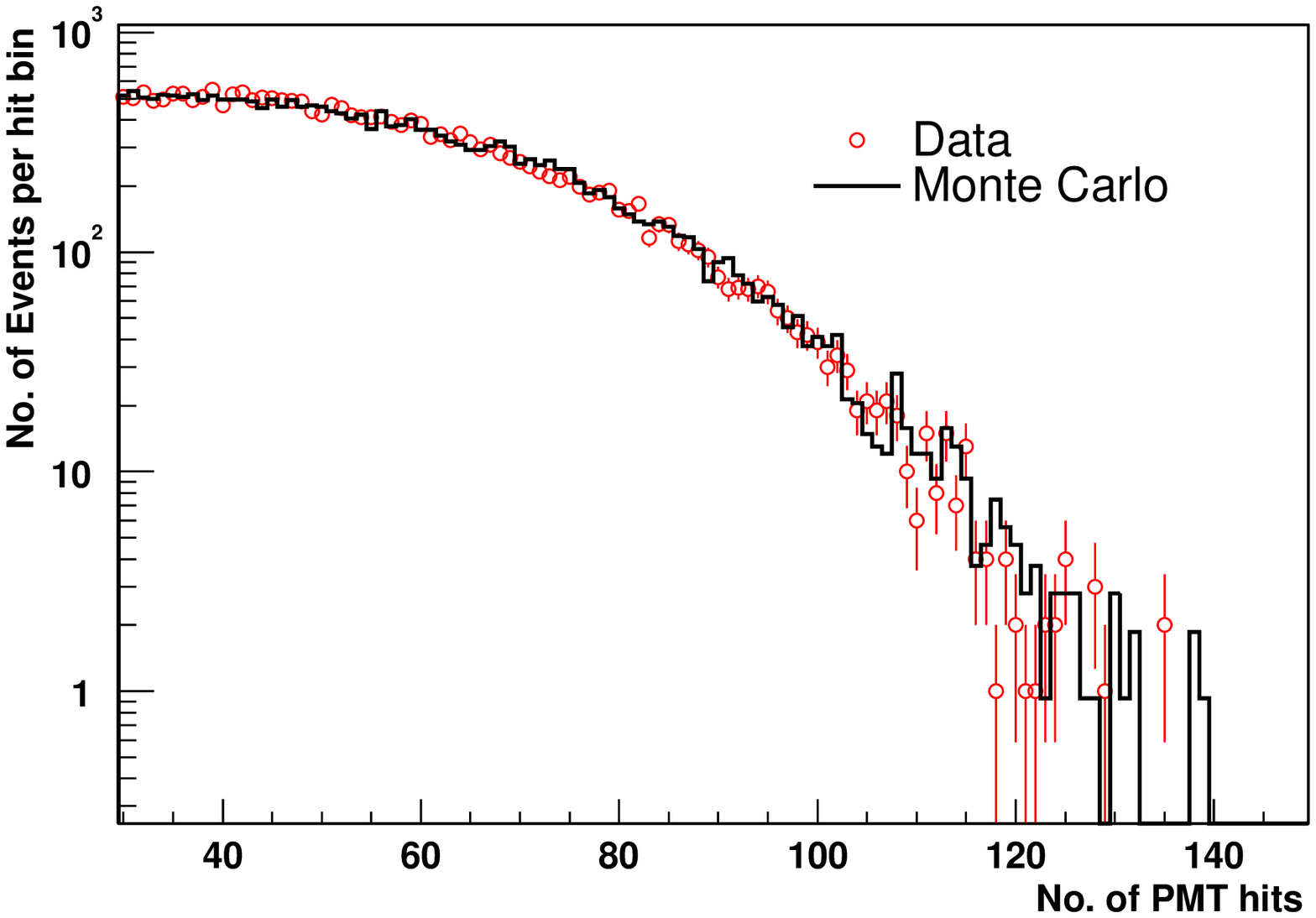}{\fLiNhitCaption}}

\newcommand{\fFitterCaption}{
The left plot shows the x-z projection of reconstructed positions of
$\sim$15k events for two placements of the $^8$Li source in the SNO
detector, (x=-16~cm,y=24~cm,z=0~cm) and (x=-16~cm,y=24~cm,z=450~cm).
The right plot shows a histogram of the distance between the source
center and the reconstructed event positions for the z=0~cm position
(solid circles) and the z=+450~cm position (open circles). (The center
of the acrylic vessel is defined to be at (0,0,0) and the z-axis is
vertically upwards.) 
}
\newcommand{\fFitter}{\myfig{f:fitter}{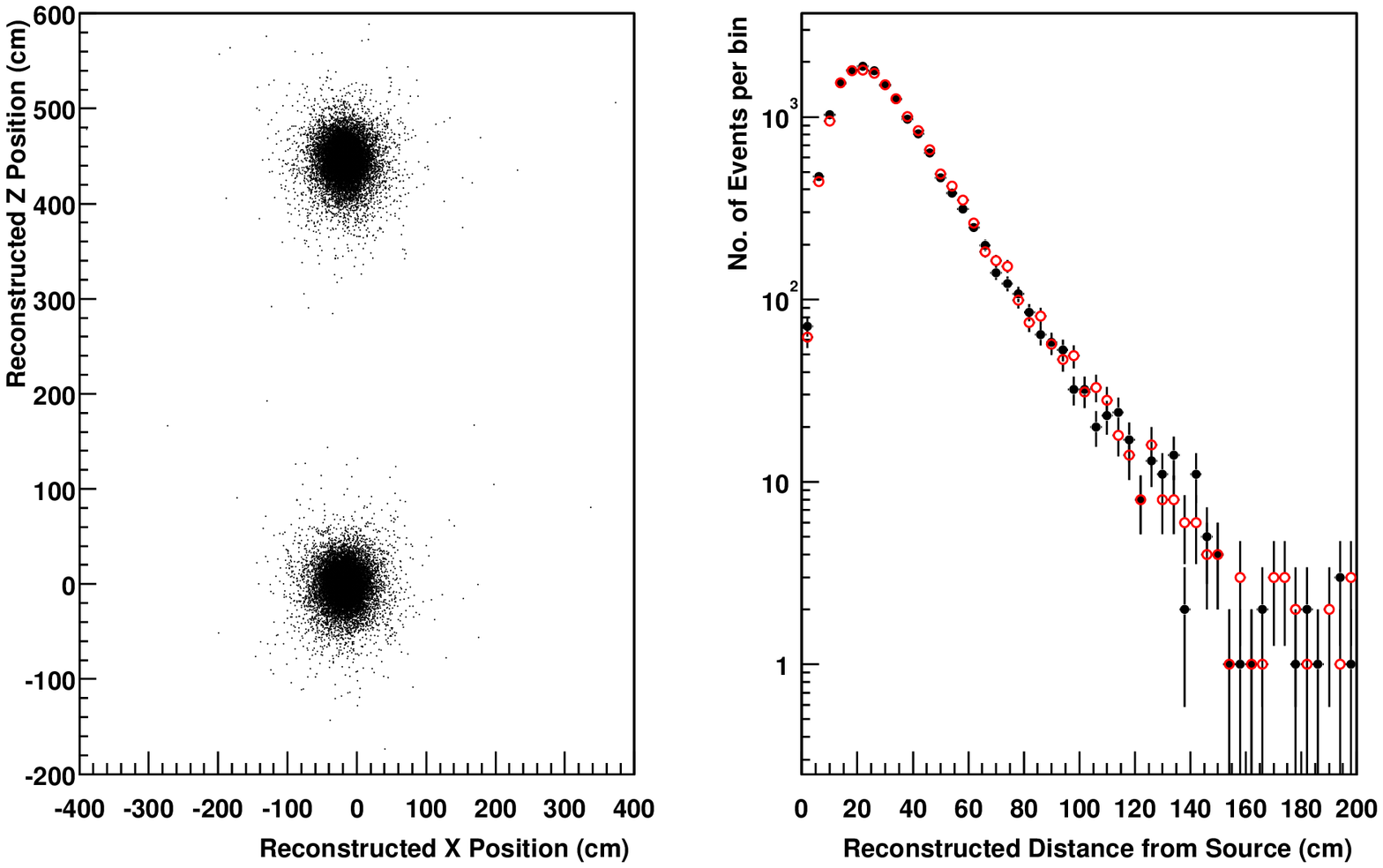}{\fFitterCaption}}

\begin{document}

\begin{frontmatter}



\title{The $^8$Li Calibration Source For The Sudbury Neutrino Observatory}

\author[Guelph]{N.J. Tagg \thanksref{Oxford}},
\author[lanl,Queens]{A. Hamer},
\author[aecl]{B. Sur},
\author[aecl,Queens]{E. D. Earle},
\author[triumf]{R.L. Helmer},
\author[aecl]{G. Jonkmans \thanksref{guy-current}},
\author[Queens]{B.A. Moffat},
\author[Guelph]{J.J. Simpson}

\address[Guelph]{University of Guelph, Guelph, Ontario N1G 2W1, Canada}
\address[lanl]{Los Alamos National Laboratory, Los Alamos, NM 87545, USA}
\address[Queens]{Queen's University, Kingston, Ontario K7L 3N6, Canada}
\address[aecl]{Atomic Energy of Canada Limited, Chalk River Laboratories, Chalk River, Ontario K0J 1J0, Canada}
\address[triumf]{TRIUMF, 4004 Wesbrook Mall, Vancouver, British Columbia, V6T 2A3, Canada}

\thanks[Oxford]{Current Address: Nuclear and Astrophysics Laboratory, Oxford University, Oxford, OX1 3RH, UK}
\thanks[guy-current]{Current Address: Bubble Technology Industries Inc., Chalk River,
Ontario, Canada K0J 1J0}

\begin{abstract}


A calibration source employing $^8$Li (t$_{1/2}$ = 0.838~s) has
been developed for use with the Sudbury Neutrino Observatory (SNO).
This source creates a spectrum of beta particles with an energy range
similar to that of the SNO $^8$B solar neutrino signal. The source is
used to test the SNO detector's energy response, position reconstruction and data
reduction algorithms.  The $^8$Li isotope is created using a
deuterium-tritium neutron generator in conjunction with a $^{11}$B
target, and is carried to a decay chamber using a gas/aerosol
transport system.  The decay chamber detects prompt alpha particles
by gas scintillation in coincidence with the beta particles which exit
through a thin stainless steel wall. A description is given of the
production, transport, and tagging techniques along with a discussion
of the performance and application of the source.

\end{abstract}

\end{frontmatter}

\section{Introduction}  \label{intro}

The Sudbury Neutrino Observatory (SNO) is a 1 kilotonne mass heavy
water Cherenkov detector designed for detection of neutrinos from the
$^{8}$B reaction in the Sun. Neutrinos interact in D$_2$O by elastic
scattering on electrons (ES), and the charged current (CC) and neutral
current (NC) interactions on deuterons.  The ES and CC interactions
both result in Cherenkov light from an energetic electron, and the NC
interaction results in the disintegration of the deuteron which is
observed by $\gamma$-rays from the capture of the free neutron.
Cherenkov light from these interactions is detected by a spherical
array of $\sim$ 9500 photomultiplier tubes surrounding the D$_2$O
volume, which is contained in a 12 meter diameter acrylic vessel.  The
detector is immersed in $\sim$7 kilotonnes of light water to provide
support and shielding.  The detector is located in a cavity in INCO's
Creighton mine near Sudbury, Canada.  A general description of the SNO
detector and its subsystems can be found in Reference
\cite{sno_nim}.

Cherenkov light measurements in SNO are based upon coincident
phototube signals registered within an event (i.e. within 400~ns).
Energy information is largely associated with the number of tubes hit
(N$_{hit}$), and the event position can be reconstructed using the
relative phototube timings within the event.  The hit pattern can be
used to determine direction information and discriminate against
instrumental backgrounds.  Use of these signals requires an
understanding of the detector response both through the use of Monte
Carlo techniques, and through analysis of calibration sources (also
discussed in Reference \cite{sno_nim}).

The SNO detector is calibrated using a variety of sources, including
an isotropic light source, $\gamma$-rays, neutrons, and
$\beta$-particles.  This work describes a source which provides a
high-endpoint energy spectrum of $\beta$-particles created by the decay of
$^8$Li. The source can be used to assist in energy calibrations, position
reconstruction studies, and evaluation of instrumental background
discrimination.

\ifpub \else \fToidecay \fi

The $^8$Li isotope is of interest for several reasons. The isotope
decays with a Q-value of 16.003~MeV and the main beta-decay branch is
an allowed decay with a central end-point energy of 12.96~MeV \cite{toi}. The
energy spectrum produced is similar to that of the $^8$B
charged-current neutrino signal, \cite{wilkinson,jonkmans-8li} both in
endpoint and shape, as $^8$Li $\beta^-$-decay is the isospin mirror to
the $^8$B $\beta^+$-decay (see Figure \ref{f:toidecay}). Both isotopes
decay to the same broad excited state of $^8$Be which in turn promptly
decays to $\alpha$-particles that can be used to identify (``tag'')
an event in the $^8$Li source.

\ifpub \else \fOverview \fi

The $^8$Li calibration source \cite{nathaniel_thesis} consists of
several subsystems, illustrated in Figure \ref{f:overview} and
described in this paper. Radioisotope production and transport are
described in sections \ref{prod} and \ref{trans}. (Proof-of-principle
and design tests of these mechanisms have been reported
seperately \cite{str:sur-li-95,*str:li_first_run}). The design
of the decay chamber is described in section \ref{decay}.  The decay
chamber is positioned using the SNO manipulator system, described in
Reference \cite{sno_nim}. The performance of the source and its application are
discussed in section \ref{perform}.

\section{$^8$Li Production} \label{prod}

The $^8$Li isotope is produced by the $^{11}$B(n,$\alpha$) reaction
(Q=-6.63 MeV, $\sigma$=30.5~mb \cite{11bxsection}) using a
deuterium-tritium (DT) generator to provide approximately 10$^{8}$
14-MeV neutrons per second.  This DT generator (MF Physics model
A-320) and its environment are described more thoroughly in Reference
\cite{n16nim}.  The primary production target consists of $>99\%$
enriched $^{11}$B boron powder suspended in a polyethylene matrix. The
target lines the inner wall of an annular target chamber that
surrounds the DT generator (see Figure \ref{f:target_chamber}). The
outer wall of this chamber is lined with a polyethylene collar whose
inner surface is coated with boron nitride.

\ifpub \else \fTargetChamber \fi

The $^8$Li atoms recoil from the target into a laminar gas stream
moving through the chamber.  Helium is used as the transport
gas, with sodium chloride aerosol added to provide an affinity for the
charged lithium atoms.  The aerosol is introduced to the gas stream
using a tube oven containing a quartz tube with NaCl heated to
approximately 610$^\circ$C, just below its melting point.

The chamber dimensions have been optimized to allow the recoiling
$^8$Li atoms just enough room to stop in the gas before hitting a wall, but
minimize the gas volume in the chamber so that the residence time of
the gas is short.  Because $^{8}$Li recoils in the direction of the incident
neutrons, it is primarily the inner target which is responsible for
$^8$Li production.

Although the total rate of isotope production in the target can be
readily calculated, the recoil and uptake by the salt aerosol are not
accurately predicted. Direct measurement of the collection rate is
complicated because corrections based on the gas transport efficiency
(see following section) are needed.  Because of the many assumptions
made in the gas transport model, it is difficult to separate $^{8}$Li
collection from transfer efficiency when measuring yields. If the gas
model described below is assumed correct then the $^8$Li recoil and
capture rates (i.e. collection rate) can be estimated by correcting
yield data with measurements from different transfer
configurations. The estimated rate at which $^8$Li atoms are
collected is (3$\pm$1) $\times 10^{-7}$ $^{8}$Li atoms per 14-MeV
neutron.

\section{Gas Transport System} \label{trans}

The isotope is transported from the shielded neutron generator pit
to the detector by two sets of transfer lines. One set, a conduit
approximately 40 meters long, connects the production chamber in
the pit to the source manipulator system on a deck above the SNO
detector.  The second set is an umbilical approximately 26 meters
long that runs through an umbilical retrieval mechanism, and down into
a decay chamber immersed in the D$_2$O.  This umbilical length is
sufficient to deploy the chamber in most of the top half of the D$_2$O
volume.

The construction of the umbilical is similar to that used for the SNO
$^{16}$N source \cite{n16nim} but with larger overall dimensions.  The
design consists of a 3.38~mm inner diameter PTFE capillary used to
flow gas into the decay chamber, and an 8.0~mm inner diameter
polyethylene tube that is used to exhaust the gas from the decay
chamber.  The inlet capillary is inside the outlet tube.  A signal
line and low-voltage DC power lines are wrapped in a helix around the
outer tube, and the whole assembly is covered with a clean,
flexible silicone sheath acting as a water seal.  Silicone potting gel
is used to seal the space between sheath and outlet tube.  This design
allows robust manipulation of the umbilical and protects the
radioactively pure D$_2$O against contamination.

Because of the low production rate and short half-life of the $^8$Li
isotope, the characteristics of the gas system (i.e. gas pressures,
flow rates, capillary dimensions) were designed for optimization of
the $^8$Li yield. The gas flow was designed to operate at the limit of
laminar flow through the conduit and umbilical.  Faster delivery of
the radioisotope is possible through the use of near-supersonic
gas flow speeds, but is deemed impractical in this case due to
considerations of safety, complications such as salt buildup inside
the decay chamber, and the difficulties of high pressure engineering.

The yield of $^8$Li may be taken as the product of the rate of
radioisotope production ($R_p$), the efficiency with which the isotope
can be extracted from the production chamber ($\epsilon_p$), the
efficiency for transport ($\epsilon_l$), and the fraction of atoms
that decay while inside the decay chamber ($\epsilon_d$).  These
properties are estimated using gas transport dynamics as described
in References \cite{van_atta,bird}.

The production chamber was designed for approximately 200-250 kPa absolute
gas pressure to optimally range out $^8$Li recoils.  Although the
chamber was designed to promote laminar flow of gas, calculations were
done for the pessimistic case of full mixing. The rate of change of
the number of atoms of $^8$Li in the production chamber is described
by Equation (\ref{eq:dndt}) for the case of full mixing, 
\begin{equation} \label{eq:dndt}
\frac{dN}{dt} = R_{p} - \lambda_{8Li}N - E
\end{equation}
where $N$ is the number of $^8$Li atoms in the chamber, $R_{p}$ is the
rate of isotope pickup on the salt in the chamber, $\lambda_{8Li}$ is
the decay rate of $^8$Li (ln2/t$_{1/2}$), and E is the rate of $^8$Li atoms
flushed from the chamber, such that $E=\frac{N}{V_p}\frac{dV}{dt}$,
with $V_p$ as the volume of the production chamber. 

We define the mass flow rate as $Q=P\frac{dV}{dt}$. The mean time to
completely flush the production chamber, $\tau_p$, is determined by
this flow rate.
\begin{equation}
\tau_p = \frac{1}{\lambda_p} = V_p\left(\frac{dV}{dt}\right)^{-1} = \frac{PV_p}{Q}
\end{equation}

The steady state solution of Equation (\ref{eq:dndt}) allows us to define
the efficiency as the fraction of produced $^8$Li which escapes the
target chamber, $\epsilon_p$.
\begin{equation}
\epsilon_p =  \frac{E}{R_p} = \frac{\lambda_p}{\lambda_p+\lambda_{8Li}} = \frac{1}{1+\frac{\tau_p}{\tau_{8Li}}}
\end{equation}

The mean time for the gas to move from the production chamber through
the transfer line to the decay chamber, $\tau_l$ is:
\begin{equation} \label{eq:taul}
\tau_l = \int_0^l \frac{P(x)}{Q} A dx 
\end{equation}
where $P(x)$ is the pressure at point $x$ along the line, $A$ is the
cross-sectional area of the transfer line, and $l$ is the total
distance.  For laminar flow this is calculable using Poisseuille's law
for mass flow, Equation (\ref{eq:pois}). In this equation, $\eta$ is
the gas viscosity, $d$ is the diameter of the transfer line, and $P_p$
and $P_d$ are the pressures of the production and decay chambers respectively.
\begin{equation}
\label{eq:pois}
Q = \frac{\pi d^4 \left(P^2_p - P^2_d\right)}{256 \eta l}
\end{equation}

For given pressures $P_p$ and $P_d$, Equation (\ref{eq:taul}) can be computed to give the
mean transfer time, Equation (\ref{eq:tflow}).

\begin{equation}
\label{eq:tflow}
\tau_{l}=\frac{2 A l}{3Q} \times \frac{P_{p}^{3}-P_{d}^{3}}{P_{p}^{2}-P_{d}^{2}} 
\end{equation}

The fraction of $^{8}$Li atoms that reach the far end defines the transfer line
efficiency, Equation (\ref{eq:epsilon_l}).

\begin{equation} \label{eq:epsilon_l}
\epsilon_l = \exp\left(-\frac{\tau_l}{\tau_{8Li}}\right)
\end{equation}

The decay chamber efficiency is the complement of the production
chamber efficiency.  Maximal gas mixing is assumed (now the optimal
case) so that the efficiency is given by Equation (\ref{eq:epsilon_d}).

\begin{equation} \label{eq:epsilon_d}
\epsilon_d =  1 - \frac{\lambda_d}{\lambda_d+\lambda_{8Li}} = \frac{1}{1+\frac{\tau_{8Li}}{\tau_d}}
\end{equation}
where $\tau_d = \frac{1}{\lambda_d} = P_d V_d / Q$, given the pressure
and volume of the decay chamber as $P_d$ and $V_d$ respectively.

The net yield of the system is computed as the product of the above efficiencies.
\begin{equation}
{\rm Yield} = R_{p} \times \epsilon_p \times \epsilon_l \times \epsilon_d
\end{equation}

The system design is constrained by both practical considerations and
by the desire to maximize this yield.  To begin with, the volume of
the decay chamber was chosen in order to get the best signal (see
Section \ref{decay}).  The decay chamber pressure was determined by the
conductance of the umbilical outlet tube, which was
exhausted to laboratory atmospheric pressure, 125 kPa (absolute) at
2000~m below surface.  

Arbitrarily high gas flow rates could not be used, because the resulting
turbulent flow would not be efficient in transporting the
radioisotope. Turbulent currents could increase the time it takes for a
given particle to transit the transfer line, and could also deposit salt
aerosol on the walls of the transfer line, both of which impair
efficiency. Laminar flow occurs only when the flow rate is such that
the Reynold's number is less than a critical value of
approximately 1000 \cite{van_atta,bird}. The Reynold's number, $R_e$
is a dimensionless quantity defined as:

\begin{equation} \label{eq:reynolds}
R_e = \frac{4 m}{\pi \eta k T} \frac{Q}{d}
\end{equation}

Here, $m$ is the molecular weight of the gas, $k$ is Boltzmann's
constant, and $T$ is the absolute temperature.  Low molecular weight is
desirable, making helium a good choice for the transport gas. To
get maximum laminar flow rate for a given capillary diameter $d$,
Equation (\ref{eq:reynolds}) is solved, using the estimated Reynold's
number of 1000:

\begin{equation}\label{eq:qmax}
Q_{max}(d) = \frac{1000 d \pi kT \eta}{4 m}
\end{equation}

\ifpub \else \fEfficiencyVsDiameter \fi

Taking $Q = Q_{max}(d)$ minimizes the transport time.  Using the flow
rate $Q_{max}(d)$ as a function of $d$, the head pressure $P_p$
needed to drive the gas flow can be determined from Equation
(\ref{eq:pois}).  The efficiencies for each component of the system
are computed for a given value of $d$ in order to find the optimum
diameter, as shown in Figure \ref{f:efficiency_vs_diameter}.  This
figure illustrates that the optimal transport capillary diameter is
approximately 3.25 mm if the decay chamber is at 125 kPa(absolute), giving an
approximate total efficiency of $\sim$ 2.5\%. Higher efficiencies may
be achievable if the decay chamber pressure can be lowered using an
exhaust pump, but this has not been tested in SNO.

The final system configuration, described in Table
\ref{t:gas_stuff}, provided a workable compromise between overall efficiency and robust
operation.

\ifpub 
\nocite{CRC}
\else 
\tGasStuff 
\fi



\section{Decay Chamber Design} \label{decay}

The decay chamber is designed to fulfill three criteria: it must (a) be
capable of detecting $^8$Li decays with a high efficiency for event
identification during analysis, (b) provide the best possible electron
spectrum to the SNO detector, and (c) it must be clean,
safe, and robust to protect the ultrapure SNO detector media. These
last two criteria are somewhat in competition, since the best electron
spectrum requires a thin-walled chamber to minimize energy loss of the
escaping beta, while safety demands that the wall be thick enough to
resist buckling when the chamber is under pressure underwater (at a
maximum depth of 18 meters of D$_2$O).  The main volume of the decay
chamber is made of spun stainless steel in a sphere, which provides
excellent resistance to compressive forces, good cleanliness for
exposure to the water, and elastic properties such that a mechanical
failure will tend to deform but not rupture the container.  The sphere
has a radius of 6.35~cm, with a wall thickness of 0.06~cm, corresponding
to an electron energy loss of approximately $\sim$1~MeV for the
electron energies of interest.

\ifpub \else \fSource \fi

A diagram of the decay chamber is shown in Figure \ref{f:source}.
Above the spherical decay volume is a conical `neck' which contains the
in- and out-flow gas ports that are connected to the umbilical
(described in Section \ref{trans}). Also in this neck is a thin acrylic
window atop which sits a 1.5'' Hamamatsu R580 photomultiplier tube (PMT)
viewing the decay volume.  The PMT, voltage supply, and gas
fixtures are contained inside a sealed stainless steel endcap sitting
atop the neck.

Event identification is performed by utilizing the scintillating
properties of the helium transport gas.  The 2$\alpha$ decay
following the $^8$Li beta decay creates two alpha tracks nearly
back-to-back with a mean energy of 1.6 MeV per particle.  In the
helium gas, each of these particles has a range of about 4~cm, so a
large proportion of the alpha energy is deposited in the gas.

The scintillation signal is not large \cite{birks:theory} and consists
mostly of UV light which is outside the acceptance of the PMT photo-cathode.
Moreover, the PMT photo-cathode coverage is a small fraction of the
decay volume, so measures were taken to enhance the signal.
The inside of the sphere is painted with a reflective white titanium
oxide based paint (Bicron BC620), and has a thin coating of
tetraphenyl butadiene (TPB) which acts as a wavelength shifter.
In addition, a small admixture of N$_2$ gas ($\sim 0.1\%$) is added to the
helium gas stream to act as an additional wavelength shifter.

The scintillation light from helium is known to be prompt
\cite{birks:theory} and the TPB and N$_2$ wavelength shifters are also fast,
so the scintillation signal was fast enough to be included in
the 400~ns coincidence window in the SNO electronics.  The tagging PMT
signal pulse height was digitized and recorded in a spare SNO
electronics channel. 

Decay tagging by gas scintillation has several advantages over other
detection methods (e.g. proportional counters, ZnS scintillation).
The decay volume need not assume a specific shape, which allows
construction of a thin-walled design. (Cylindrical shapes, such as
those required for proportional counters, require a larger wall
thickness for the same volume container.)  Gas scintillation is much
faster than many other methods, allowing the decay tag to be measured
in coincidence with the associated (SNO detector) event, and making
subsequent analysis simple and robust.

\section{Source Performance And Application} \label{perform}

The $^8$Li source has been successfully tested and deployed in the SNO
detector. Calculations using the described gas model and data from
short-transfer line bench tests yield a prediction for the final
$^8$Li yield of 0.75 $\pm$ 0.25 decays per second. A yield of
approximately 0.5 decays per second was observed with the source
deployed within the D$_2$O volume of the SNO detector. An unexpected
feature of the system behaviour was that the source took $\sim$2 hours
to climb to the maximum yield. This is likely due to the salt
temperature reaching equilibrium, increasing the pickup efficiency of
the aerosol by the helium gas stream.  This equilibration was not
observed in short run bench tests.  The yield provided by the source
allowed a collection of some 50 000 tagged $^8$Li events to be
observed.

Bench tests, using coincidences of external scintillators to identify
$^8$Li electrons, allowed a measurement of the tagging efficiency of
the decay chamber.  This efficiency was estimated at 92$\pm$5\% of all
$^8$Li decays in the ball. This tagging efficiency remained quite
robust with changes in decay chamber pressure.  The tag signal was
prompt ($\sim$150~ns after the decay), sufficient to record the signal
within the 400~ns coincidence window of the SNO electronics.  The PMT
tagging signal from each $^8$Li decay was approximately 10-20 photoelectrons.

The $^8$Li system generated signals created by contaminant
radioisotopes.  Evidence of these contaminants was indicated in bench
tests, and was confirmed during deployment in the SNO detector.  In
addition to 0.5~$s^{-1}$  $^8$Li decays, approximately 8~$s^{-1}$ of
$\sim$5~MeV events were observed.  Analysis of the signal indicated
that the majority of this contamination was likely to be $^{16}$N, an
isotope that decays by beta-gamma decay with a 6.13~MeV gamma and a
half-life of 7.13~s. This isotope was likely produced by neutron
reactions such as $^{19}$F~(n,$\alpha$) and $^{16}$O~(n,p) on Teflon
and plastic components in the target chamber, and then carried to the
decay chamber by the gas stream.

This contamination signal, however, did not interfere with the $^8$Li
signal, because the majority of these events generated no tag pulse.
The beta-gamma contaminants have only a small chance ($\sim$0.6\%) for
the beta particle to interact with the acrylic window at the top of
the decay volume, generating 1-2 photoelectrons in the PMT by 
Cherenkov radiation. (The gas scintillation signal created by gammas or
betas is much too small to be detected.) This contamination was
reduced even further by using the time and integrated charge of the
tagging signal to discriminate against the contaminant events.  The
PMT anode pulses are much smaller than those due to alpha particles
from $^8$Li decays, allowing a clear separation of signals.  In
addition, the PMT timing is different (i.e. has a faster rise time)
for $\beta$ detection in the chamber.  These properties were tested by
running CO$_2$ gas through the system, generating large quantities of
$^{16}$N isotope directly (in a manner completely identical with that
described in Reference \cite{n16nim}).  This test showed that the
decay chamber detected $^{16}$N in a fashion consistent with the
observed contaminant activity.

\ifpub \else \fTagSignal \fi

Event discrimination is illustrated in Figure \ref{f:tag_signal}, a
scatter plot of PMT time and charge for the $^{8}$Li runs in the SNO
detector. The $\beta$-tagged contaminant events are clearly
distinguishable from the $^{8}$Li $\alpha$ signals. Figure
\ref{f:raw_nhit} shows the N$_{hit}$ distribution for position
reconstructed events of all $^{8}$Li runs. It also shows N$_{hit}$
distributions for the remaining events with a tag requirement and
additional software cuts on PMT charges and times to distinguish
$^{8}$Li from $\beta$-tagged contaminant events such as
$^{16}$N. Using event discrimination, we estimate that the total
probability of a single contaminant event to be misidentified as an $^8$Li
event to be less than 0.005\%.  With the observed event rates, we
conservatively estimate that events due to contaminants make up less
than 0.1\% of the final $^8$Li data set.

\ifpub \else \fRawNhit \fi

\ifpub \else \fLiNhit \fi

The N$_{hit}$ response of the SNO detector to tagged $^8$Li decays is
shown in Figure \ref{f:li_nhit} for a run with the source situated at
the center of the SNO detector.  The N$_{hit}$ distribution is shown
to agree well with a simulation of the source using the SNO Monte
Carlo code and an energy scale established with the $^{16}$N source
\cite{n16nim,sno:prl}.  This comparison forms one of the suite of
calibrations used to test energy response in the full SNO analyses
\cite{sno:prl}.  Important systematic uncertainties associated
with such a comparison are uncertainties in the $\beta$ energy loss
through the stainless steel wall and optical properties (shadowing and
reflectivity) of the chamber. These uncertainties are on the order of
2\% for the energy scale, with further investigation underway. An
intriguing application of these data is direct comparison with the observed SNO
$^8$B neutrino spectrum. Such a comparison has significant potential
\cite{jonkmans-8li} to reduce systematic uncertainties associated with
$^8$Be states common to both the $^8$B and $^8$Li decay channels.

\ifpub \else \fFitter \fi

The source has been deployed in multiple locations within the SNO
D$_2$O volume, and the resulting signals have been reconstructed in
the same manner as neutrino data.  A sample of reconstructed event
positions is shown in Figure \ref{f:fitter}.  Comparison of the
source position and reconstructed event position has been used to
evaluate the accuracy and resolution of position reconstruction
algorithms \cite{nathaniel_thesis,mark_thesis}.
%

Finally, the $^8$Li data comprise one of the tests used to evaluate
data reduction techniques employed to remove instrumental background
from the SNO detector data.  The set of tagged $^8$Li events is used
as a representative sample of data with distributions nearly identical
to that of neutrino interactions in the detector.  By counting the
fraction of $^8$Li events removed by the data reduction cuts, the
number of neutrino events falsely sacrificed by the cuts may be
estimated, as described in Reference \cite{sno:prl}.

\section{Conclusions} \label{concl}

The $^8$Li source developed for SNO provides several necessary
functions required to test the SNO detector response.  Despite
difficulties in producing and delivering the short-lived isotope,
careful optimization allows an appreciable quantity of $^8$Li to be
safely delivered over moderate distances.

Helium scintillation provides a mechanism to tag $^8$Li decays in a
robust fashion. This method provides a signal that is sufficiently
prompt to do fast coincidence with other signals, and has a high tagging
efficiency.  This signal can be used to discriminate against
beta-gamma radioactive contaminants produced by the neutron source.

The source has been used successfully to evaluate SNO detector
responses for energy, vertex reconstruction, and event topology \cite{sno:prl}.
These calibrations are an important part of the SNO calibration
programme \cite{sno_nim}.

\section{Acknowledgements}

The authors wish to thank R. Deal, P. Dmytrenko, J. Fox, E. Gaudette,
V. Koslowsky, and E. Hagberg for their expertise and assistance in
construction of the target and transfer systems, K. McFarlane,
P. Liimatainen and D. Sinclair for their hard work in construction of
the $^8$Li decay chamber, and C. Hearns and F. Duncan, for their work
in deployment.  We thank P. Huffman for kindly providing the TPB
wavelength shifter. We are indebted to J.R. Leslie for suggesting the
$^{11}$B(n,$\alpha$) reaction as a production method, and E. Bonvin
for first pointing out the possiblilites of using $^8$Li as a
calibration source.  Many thanks go to the entire SNO collaboration
for their support and help in making this project possible.

The SNO project has been supported in Canada by the Natural Sciences
and Engineering Research Council, Industry Canada, National Research
Council of Canada, Northern Ontario Heritage Fund Corporation and the
Province of Ontario, in the United States by the Department of Energy,
and in the United Kingdom by the Science and Engineering Research
Council and the Particle Physics and Astronomy Research Council. The
heavy water has been loaned by AECL with the cooperation of Ontario
Power Generation.  The provision by INCO of an underground site is
greatly appreciated.


\ifpub

\clearpage
Table \ref{t:gas_stuff}: \tGasStuffCaption
\tGasStuff

\clearpage
Figure \ref{f:toidecay}: \fToidecayCaption

Figure \ref{f:overview}: \fOverviewCaption

Figure \ref{f:target_chamber}: \fTargetChamberCaption

Figure \ref{f:efficiency_vs_diameter}: \fEfficiencyVsDiameterCaption

Figure \ref{f:source}: \fSourceCaption

Figure \ref{f:tag_signal}: \fTagSignalCaption

Figure \ref{f:raw_nhit}: \fRawNhitCaption

Figure \ref{f:li_nhit}: \fLiNhitCaption

Figure \ref{f:fitter}: \fFitterCaption
\clearpage

\fToidecay
\fOverview
\fTargetChamber
\fEfficiencyVsDiameter
\fSource
\fTagSignal
\fRawNhit
\fLiNhit
\fFitter

\fi

\bibliographystyle{elsart-num.bst}
\bibliography{lipaper}

\end{document}